\documentclass[12pt]{article}
\usepackage{amsmath}
\usepackage{graphicx}
\usepackage{amssymb} 
\newcommand{\Mat}[1]{{{\boldsymbol{#1}}}}

\def\be{\begin{equation}}
\def\ee{\end{equation}}
\def\bea{\begin{eqnarray}}
\def\eea{\end{eqnarray}}
\def\dd{\mathrm{d}}

\date{29 June 2007}
\begin{document}

\title{Quantum wave equations in curved space-time\\ from wave mechanics}

\author{
Mayeul Arminjon\\
\small\it Laboratoire ``Sols, Solides, Structures, Risques'' (CNRS \& Universit\'es de Grenoble)\\
\small\it BP 53, F-38041 Grenoble cedex 9, France.
\footnote {\,Part of this work was done while the author was at Dipartimento di Fisica, Universit\`a di Bari and INFN Bari, Italy.} 
}

\maketitle
\begin{abstract}
Alternative versions of the Klein-Gordon and Dirac equations in a curved spacetime are got by applying directly the classical-quantum correspondence.\\
\end{abstract}
\hspace{5mm}{\bf Key words:} quantum mechanics in a gravitational field; Dirac equation.\\
\vspace{5mm}

The usual way to write the wave equations of relativistic quantum mechanics in a curved spacetime is by {\it covariantization}: the searched equation in curved spacetime should coincide with the flat-spacetime version in coordinates where the connection cancels at the event $X$ considered. This is connected with the {\it equivalence principle.} For the Dirac equation with standard (spinor) transformation, this procedure leads to the Dirac-Fock-Weyl (DFW) equation, which does {\it not} obey the equivalence principle. Alternatively, in this work we want to {\it apply directly the classical-quantum correspondence.}\\

The latter results \cite{A22} from two mathematical facts \cite{Whitham}. {\bf i}) There is a one-to-one correspondence between a (2$^\mathrm{nd}$-order, say) linear differential operator:
\be\label{wave-eqn}
\mathrm{P}\psi \equiv \ a_0(X) +a_1^\mu (X)\partial _\mu \psi +a_2^{\mu \nu } (X)\partial _\mu \partial _\nu\psi,
\ee
and its {\it dispersion equation,} a polynomial equation for covector ${\bf K}$:
\be\label{dispersion-eqn} 
\Pi_X({\bf K})\equiv a_0(X) +i\,a_1^\mu (X)K_\mu +i^2a_2^{\mu \nu } (X)K_\mu K_\nu  =0, 
\ee
the latter arising when one looks for ``locally plane-wave" solutions \cite{A22}: $\ \psi (X) = A\,\exp[ i\theta (X)]$, with $\ \partial_\nu K_\mu (X_0)=0$, where $ K_\mu \equiv \partial _\mu \theta$. The correspondence from (\ref{dispersion-eqn}) to (\ref{wave-eqn}) is
$K_\mu  \rightarrow \partial _\mu/i$. {\bf ii}) The propagation of the spatial wave covector $\ {\bf k}\equiv (K_j)\ (j=1,2,3)$ obeys a {\it Hamiltonian system:}
\be \label{Hamilton-W-k} 
\frac{\dd K_j}{\dd t}= -\frac{\partial  W}{\partial x^j},\qquad \frac{\dd x^j}{\dd t}= \frac{\partial  W}{\partial K_j}\qquad (j=1,2,3),
\ee
where $W({\bf k};X)$ is the {\it dispersion relation}, got by solving $\, \Pi_X({\bf K}) =0$ for the frequency $\,\omega\equiv -K_0$. Wave mechanics (classical trajectories=skeleton of a wave pattern) means that the classical Hamiltonian is $H=\hbar W$. The classical-quantum correspondence follows \cite{A22} by substituting $K_\mu  \rightarrow \partial _\mu/i$.\\

This analysis shows \cite{A39} that {\it the classical-quantum correspondence needs using preferred classes of coordinate systems:} the dispersion polynomial $\Pi _X({\bf K})$ and the condition $\partial_\nu K_\mu (X)=0$ stay invariant only inside any class of ``infinitesimally-linear" coordinate systems, connected by changes satisfying, at point $X$ considered, $\frac{\partial ^2x'^\rho}{\partial x^\mu\partial x^\nu}=0, \ \mu ,\nu ,\rho  \in \{0,...,3\}.$ One such class is that of locally-geodesic coordinate systems at $X$ for metric $\Mat{g}$: $g_{\mu \nu,\rho}(X)=0, \ \mu,\nu, \rho \in \{0,...,3\}.$ Another class occurs if there is a (physically) preferred reference frame: that made of changes which are internal to this frame. Assuming one class or the other gives distinct wave equations.\\

We may now apply this correspondence to a relativistic particle, {\it also in a curved space-time} \cite{A39}. In each coordinate system, the energy component $p_0$ of the 4-momentum defines a classical Hamiltonian $H\equiv -p_0$ satisfying
\be \label{relativistic-energy-eqn}
g^{\mu \nu} p_\mu p_\nu -m^2 =0 \qquad (c=1).
\ee
The dispersion equation associated with this by wave mechanics is
\be \label{Relativistic-dispersion}
g^{\mu \nu} K_\mu K_\nu -m^2= 0 \qquad (\hbar=c=1).
\ee
Applying directly the correspondence $K_\mu  \rightarrow \partial _\mu/i$ to it, leads to the Klein-Gordon equation. Instead, one may try a {\it factorization:}
\be \label{Pi-factorization}
\Pi_X({\bf K})\equiv (g^{\mu \nu}(X) K_\mu K_\nu -m^2){\bf 1}= [\alpha(X) +i \gamma ^\mu(X) K_\mu][\beta(X) +i \zeta^\nu (X) K_\nu].
\ee
Identifying coefficients in (\ref{Pi-factorization}) (with noncommutative algebra), and then substituting $K_\mu  \rightarrow \partial _\mu/i$, leads to the Dirac equation:
\be \label{Dirac-brut}
(i\gamma^\mu \, \partial _\mu -m) \psi=0,\qquad \mathrm{with\ }\gamma ^\mu \gamma ^\nu + \gamma ^\nu \gamma ^\mu = 2g^{\mu \nu}\,{\bf 1}.
\ee

Assume the first class (locally-geodesic systems). Then Eq. (\ref{Dirac-brut}), derived in any system of that class, rewrites in a {\it general} coordinate system as:
\be \label{Dirac-general-equivalence principle}
\left(i\gamma^\nu D_\nu -m\right)\psi =0,\qquad (D_\nu \psi )^\mu \equiv \psi ^\mu_{;\,\nu} \equiv \partial _{\nu}\psi^\mu + \Gamma ^\mu_{\sigma \nu }\psi^\sigma.
\ee  
(The $\Gamma ^\mu_{\sigma \nu }$'s are the Christoffel symbols of $\Mat{g}$.) With the second class (preferred-frame systems), a different (preferred-frame) equation is got. These two equations are also distinct from the standard, DFW equation.

\bibliographystyle{ws-procs9x6}

\end{document}